\documentclass[10pt,aps,twocolumn, prl, footinbib]{revtex4-1}
\usepackage{graphicx}
\usepackage{psfrag}
\usepackage{amsmath,amssymb}
\usepackage{colordvi}
\usepackage{color}
\usepackage{hyperref}
\hypersetup{
    colorlinks=true, 
    linktoc=all,     
    linkcolor=blue,  
    urlcolor=blue,
    citecolor=red,
}

\begin{document}

\title{Normal State Properties of a Resonantly Interacting $p$-wave Fermi Gas}
\author{Juan Yao}
\affiliation{Department of Physics and Centre of Theoretical and Computational Physics, The University of Hong Kong, Hong Kong, China}

\author{Shizhong Zhang}
\email[]{shizhong@hku.hk}
\affiliation{Department of Physics and Centre of Theoretical and Computational Physics, The University of Hong Kong, Hong Kong, China}
\date{\today}

\begin{abstract}
Motivated by the recent experimental progress in the study of $p$-wave resonant Fermi gases, we investigate the normal state properties of such a gas. We calculate the universal equation of state and the two $p$-wave contacts that characterise the universal properties of the system, in good agreement with experiments. Our calculation takes explicit account of the effective range correction, obtains the superfluid transition temperature $T_c$ within the Nozi{\`e}res-Schmitt-Rink (NSR) scheme, and shows that it lies within experimental reach. We derive an analytic expression for $T_c$ in the weak coupling limit and show explicitly the non-perturbative nature of the effective range corrections.
\end{abstract}
\maketitle

{\em Introduction.} Studies of $p$-wave Feshbach resonances with ultracold atomic gases date back to 2003 when they were first observed in $^{40}$K~\cite{Regal2003}. This and subsequent experiments~\cite{Zhang2004,Gunter2005,Schunck2005,Gaebler2007,Fuchs2008,Inada2008,Nakasuji2013,Ticknor2004,Chevy2005,Waseem2016} have explored the collisional properties of the gas in detail and showed that the system suffers significant loss close to resonance. In contrast, broad $s$-wave Feshbach resonances are stabilized by Pauli exclusion of three-body processes, allowing an s-wave resonant superfluid to be realized~\cite{Zwerger2011}. The difference between the $s$- and $p$-wave resonances stems from the existence of centrifugal barrier in the case of $p$-wave resonance, which tends to support quasi-bound dimer states within the centrifugal barrier and thus leads to significant atom loss, preventing the study of a resonant $p$-wave gas in equilibrium~\cite{Gaebler2007,Levinsen2008}. 

However, a recent experimental study~\cite{Luciuk2016} of a single component Fermi gas of $^{40}$K, utilizing a fast spectroscopic measurement, has shown that close to the $p$-wave Feshbach resonance, the system can establish quasi-equilibrium between the scattering fermions and the quasi-bound dimer states, while suffering an overall loss that still allows the study of properties of the gases to be conducted. In this way, it is demonstrated experimentally that the $p$-wave resonant Fermi gas obeys a set of universal relations controlled by the $p$-wave contacts~\cite{Yoshida2015,Yu2015,He2016,Peng2016,Cui2016}, analogous to the $s$-wave case~\cite{Tan2008,Braaten2008,Zhang2009,Werner2009}. Unlike the $s$-wave case where usually a single parameter, the $s$-wave scattering length, is sufficient for the description of the universal properties, in the $p$-wave case, one needs to take into account the effective range corrections in order to formulate a consistent theory~\cite{Yu2015}. In the presence of externally or spontaneous broken axial rotation symmetry, the $p$-wave contacts have to be extended to a tensor~\cite{Zhang2016,Yoshida2016} (see also Ref.~\cite{Qi2016}).  So far, however, no explicit calculation of the $p$-contacts exists except via Virial expansion~\cite{Yu2015}, leaving unexplored an exciting regime of $p$-wave resonant Fermi gases in the normal state, and in particular, close to the superfluid transition temperature. 

In this Letter, we fill this gap by developing a many-body theory for the normal state of a single component $p$-wave Fermi gas. We adopt a two-channel formulation of $p$-wave resonances and extend the Nozi{\`e}res-Schmitt-Rink (NSR) scheme for $s$-wave resonances~\cite{2NSR}, emphasizing the role of $p$-wave effective range. We calculate the universal equation of state for the resonant Fermi gases in the normal state and also the associated $p$-wave contacts. Finally we determine the superfluid transition temperature $T_c$ of a $p$-wave Fermi gas, using parameters appropriate to the current experiment. An analytic expression for $T_c$ is also obtained in the weak coupling limit that shows explicitly its non-perturbative dependence on the effective range, and further emphasizes its special importance as compared with $s$-wave case.

{\em The Model}. We adopt a two-channel description of the $p$-wave Feshbach resonance for a spinless Fermi gas. The non-interacting Hamiltonian:
\begin{equation}
\hat{H}_0=\sum_{{\bf k}}\epsilon_{{\bf k}}\hat{a}_{{\bf k}}^\dagger\hat{a}_{{\bf k}}
+\sum_{m,{\bf q}}(\epsilon_{{\bf q}}/2-\nu_{m})\hat{b}_{m,{\bf q}}^\dagger\hat{b}_{m,{\bf q}},
\end{equation}
where $\hat{a}_{{\bf k}}^\dagger$ is the creation operator for (spinless) fermions with momentum $\hbar{\bf k}$ with kinetic energy $\epsilon_{{\bf k}}=\hbar^2{\bf k}^2/2M$, where $M$ is the fermion mass. $\hat{b}_{m,{\bf q}}^\dagger$ is the creation operator for closed channel molecules with momentum $\hbar{\bf q}$ and relative angular momentum projection $m$. In this work, we work close to a $p$-wave resonance and neglect other partial wave scatterings, including the $s$-wave. As a result, $m=0,\pm 1$. In actual experiment for $^{40}$K, the resonances for $m=0$ and $m=\pm 1$ are split by about $0.5$G~\cite{Regal2003,Luciuk2016}. This is taken into account by assuming a $m$-dependent detuning $\nu_m$ of the closed channel molecules. The conversion between the open channel scattering fermions and the closed channel molecules is given by 
\begin{equation}
\hat{V}=\sum_{m, {\bf k},{\bf q}} \frac{g_m}{\sqrt{2V}}kY_{1m}(\hat{{\bf k}})\hat{a}^\dagger_{\frac{{\bf q}}{2}-{\bf k}}\hat{a}^\dagger_{\frac{{\bf q}}{2}+{\bf k}}\hat{b}_{m,{\bf q}}+{\rm H.c.},
\end{equation}
where $g_m$ is the coupling constant. The matrix element $kY_{1m}(\hat{{\bf k}})$ arises from the $p$-wave symmetry of the molecules where $k=|{\bf k}|$ and $\hat{\bf k}={\bf k}/k$. The bare coupling constants $\{\nu_m, g_m\}$ can be related to the low-energy $p$-wave scattering parameters by a standard renormalisation procedure (hereafter we set $\hbar=1$)~\cite{Yoshida2015,Luciuk2016}
\begin{align}
\label{EqC5Renor1}
v_m^{-1}&=\frac{4\pi\nu_m}{g_m^2M}+\frac{4\pi}{V}\sum_{{\bf k}}1,\\
\label{EqC5Renor2}
R_m^{-1}&=\frac{4\pi}{g_m^2M^2}+\frac{2\pi}{MV}\sum_{{\bf k}}\frac{1}{\epsilon_{{\bf k}}},
\end{align}
where we have introduced the $p$-wave scattering volume $v_m$ and effective range $R_m$ through the low-energy effective range expansion $k^3\cot\delta_m(k)=-1/v_m-k^2/R_m$. $\delta_m(k)$ is the $p$-wave scattering phase shift with magnetic projection $m$. Note that we have neglected the direct $p$-wave interaction between the spin polarized fermions.  The total number of fermions $N$ is given by $N=N_{\rm F}+2N_{\rm B}\equiv N_{\rm F}+2\sum_mN_{{\rm B},m}$, where $N_{\rm F}=\sum_{\bf k}a^\dagger_{\bf k}a_{\bf k}$ and $N_{{\rm B},m}=\sum_{\bf k}b^\dagger_{m,{\bf k}}b_{m,{\bf k}}$. In the following, we define the Fermi wave vector $k_F$ by $n\equiv N/V=k_F^3/(6\pi^2)$, where $n$ is the number density and $V$ is the volume of the system.

\begin{figure}[ht]
\begin{centering}
\includegraphics[width=0.4\textwidth]{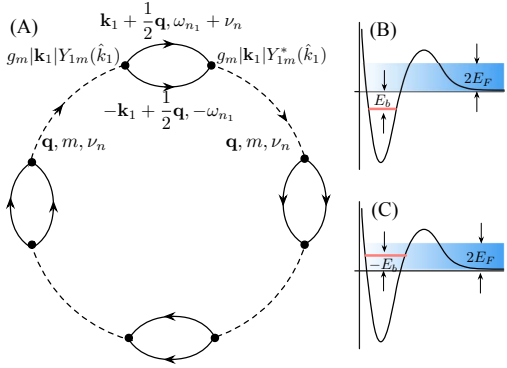}
\caption{(Color online) (A) Diagrams that contribute to the thermodynamic potential in the Nozi{\`e}res-Schmitt-Rink scheme. The solid lines represent the Green's functions for fermions $G^{\rm A}_0({\bf k},i\omega_n)=(i\omega_n-(\epsilon_k-\mu))^{-1}$. The dashed lines represent the Green's function for molecules $G^{\rm M}_0({\bf q},i\nu_n)$ of Eq.\eqref{EqG0M}. The vertex is given by $g_m|{\bf k}_1|Y_{1m}(\hat{k}_1)$, indicating the $p$-wave scattering through channel $m$. (B) Schematic diagrams of an actual bound state ($E_b>0$) that is below the scattering threshold when $v_m>0$. (C) For $v_m<0$, there is a quasi-bound state in the continuum with energy $-E_b$ above the threshold. The scattering energy of two fermions extends to $2E_F$ in a degenerate Fermi gas.} 
\label{Figring}
\end{centering}
\end{figure}

{\em Nozi{\`e}res-Schmitt-Rink scheme for $p$-wave gases}. In the normal state, the thermodynamic potential $\Omega$ can be written as $\Omega=\Omega_0^{\rm F}+\Omega_0^{\rm M}+\Omega_{\rm int}$, where $\Omega_0^{\rm F}=-1/\beta\sum_{\bf k}\ln[1+\exp(-\beta\xi_{\bf k})]$ gives the contribution of non-interacting fermions. $\beta=1/k_{\rm B}T$ and $\xi_{\bf k}=\epsilon_{\bf k}-\mu$ is the kinetic energy of fermions measured from its chemical potential $\mu$. $\Omega_0^{\rm M}={1}/{\beta}\sum_{m,{\bf q}}\ln(1-\exp{[-\beta(\epsilon_{{\bf q}}/2-2\mu-\nu_m)}])$ gives the contribution from bosonic molecules. Note that while $\Omega_0^{\rm F}$ depends only on physical parameters, the expression for $\Omega_0^{\rm M}$ involves the bare detuning $\nu_m$, which has to be renormalized later. Within NSR~\cite{2NSR}, the contribution to $\Omega$ from the interaction term is given by the ring diagrams in Fig.~\ref{Figring}. Explicitly, we have
\begin{align}\nonumber
\Omega_{\rm int}&=\sum_{m,{\bf q}}\int_{-\infty}^{\infty} \frac{dz}{\pi}\frac{1}{e^{\beta z}-1}\times \\
&{\rm Im}\{\ln[1+g_m^2\Pi_m({\bf q},z+i0^+)G_0^{\rm M}({\bf q},z+i0^+)]\},
\label{intOmega}
\end{align}
where $G_0^{\rm M}({\bf q},i\nu_n)$ is the Green's function for non-interacting molecules, 
\begin{equation}
G_0^{\rm M}({\bf q},i\nu_n)=\frac{1}{i\nu_n-(\epsilon_{{\bf q}}/2-\nu_m-2\mu)}
\label{EqG0M}
\end{equation}
and $i\nu_n=2\pi ni/\beta\hbar$ is the bosonic Matsubara frequency with integer $n$. The polarization $\Pi_m$ describes the propagation of two fermions and is given by
\begin{align}\nonumber
\Pi_m({\bf q},i\nu_n)= &\frac{1}{V}\sum_{\bf k} \Big\{k^2 |Y_{1m}(\hat{{\bf k}})|^2 \\
&\times\frac{1-f(\xi_{{\bf k}+{\bf q}/2})-f(\xi_{-{\bf k}+{\bf q}/2})}{\xi_{{\bf k}+{\bf q}/2}+\xi_{-{\bf k}+{\bf q}/2}-i\nu_n}\Big\},
\label{Pi}
\end{align}
where $f(\xi)=(\exp(\beta\xi)+1)^{-1}$ is the Fermi distribution function. A direct evaluation of $\Pi_m$ shows that it is divergent and requires renormalization. This can be achieved together with the renormalization of $\Omega_0^{\rm M}$ by noticing that 
\begin{equation}
\ln[1+g_m^2\Pi_mG_0^{\rm M}]=\ln g_m^2+\ln G_0^{\rm M}+\ln[g_m^{-2}(G_0^{\rm M})^{-1}+\Pi_m].
\end{equation}
The first term is a constant and can be neglected. The second term, when integrated in Eq.~\eqref{intOmega} cancels precisely $\Omega_0^{\rm M}$. One is thus left with the last term, which, by means of the renormalization conditions, Eqs.~(\ref{EqC5Renor1},\ref{EqC5Renor2}), reduces to 
 \begin{align}
 \Omega\equiv \Omega_0^{\rm F}+\widetilde{\Omega}_{\rm int}=\Omega_0^{\rm F}-\sum_{m,{\bf q}}\int \frac{d z}{\pi}\frac{1}{e^{\beta z}-1}\delta_m({\bf q},z),
 \label{Omegaint}
 \end{align}
an expression of identical form as that for the single channel model~\cite{Pethick2008}. Here $\delta_m({\bf q},z)$ is the $p$-wave scattering phase shift and is given by $\delta_m({\bf q},z)=-{\rm Arg}[\Gamma_m^{-1}({\bf q},z)]$, where the vertex function is given by
\begin{equation}
\Gamma_m^{-1}({\bf q},z)=\frac{M^2}{4\pi R_m}\bar{z}+\frac{M}{4\pi v_m}+\Pi^r_m({\bf q},z).
\end{equation}
The renormalized polarization $\Pi_m^r({\bf q},z)$ is
\begin{align}
\Pi^r_m({\bf q},z)=\Pi_m({\bf q},z)-\frac{M}{V}\sum_{\bf k}1-\frac{M\bar{z}}{2V}\sum_{\bf k}\frac{1}{\epsilon_{\bf k}},
\end{align}
where $\bar{z}=z-\epsilon_{\bf q}/2+2\mu$.

The structure of the vertex function $\Gamma_m$ be analyzed most easily at high temperature where one can neglect the Fermi distribution factors in Eq.~(\ref{Pi}). In that limit we find analytically $\Gamma_m^{-1}({\bf q},z)={M^2}/({4\pi R_m})\bar{z}+{M}/({4\pi v_m})+{M^{5/2}}/({4\pi})(-\bar{z})^{3/2}$. This leads to two bound states where $\bar{z}<0$. Solving the equation $\Gamma_m^{-1}({\bf q},z)=0$, one finds $\bar{z}_1=-R_m/(Mv_m)$ and $\bar{z}_2=-1/(MR_m^2)$. We are mostly interested in the resonant regime where the dimensionless parameter $|k_F^3v_m|\gg 1$, and correspondingly $R_m$ assumes its natural scale in low-energy scattering. For all experiments studying p-wave resonances, $k_F R_m \ll 1$. This means that $\bar{z}_2$ corresponds to a very deeply bound state which lies outside the validity of effective range expansion used in our work. In fact, it is known that $\bar{z}_2$ corresponds to the {\em ghost field} where the normalization for such a state becomes negative~\cite{Yusuke2012,Braaten2012}. As a result, in the implementation of NSR calculation, we should neglect the $\bar{z}_2$ pole of $\Gamma_m^{-1}$, which is allowed if one is only interested in the low-energy properties of the system. 

The other pole $\bar{z}_1=-R_m/(Mv_m)<0$ corresponds to a weakly bound state when $v_m>0$. This will be referred to as the BEC side of the resonance [see Figure \ref{Figring} (B)]. As one tunes across the resonance ($v_m=\pm\infty$) towards the BCS side ($v_m<0$), the bound state emerges above zero energy and becomes a quasi-bound state because of the $p$-wave centrifugal barrier [see Figure \ref{Figring} (C)]. The quasi-bound state couples strongly to scattering fermions and generates effective $p$-wave attraction between them. For $k_{\rm B}T\ll |\bar{z}_2|$, it is then crucial to take into account the contribution from $\bar{z}_1$ pole, whose energy we denote as as $E_{b,m}=R_m/(Mv_m)$ below.

In the high temperature limit $E_F\ll k_{\rm B}T\ll |\bar{z}_2|$, $\widetilde{\Omega}_{\rm int}$ can be expanded to lowest order in fugacity $\exp(\beta\mu)$. We can write $\delta_m({\bf q},z)=\delta_m^{\rm M}({\bf q},z)+\delta_m^{\rm F}({\bf q},z)$ in Eq.~(\ref{Omegaint}), where $\delta_m^{\rm M}({\bf q},z)$ arises from the molecular pole $\bar{z}_1$ and $\delta_m^{\rm F}({\bf q},z)$ for the scattering fermions. In the absence of the many-body medium effects (neglecting the Fermi distribution function in Eq.~(\ref{Pi})), one can reduce Eq.~(\ref{Omegaint}) to the standard virial expansion results where $\delta_m^{\rm M}({\bf q},z)$ gives the contribution from bound molecules and $\delta_m^{\rm F}({\bf q},z)$ for the scattering fermions~\cite{Pethick2008}. At temperature $T>T_c$ and for a given set of scattering parameters $\{v_m,R_m\}$, one can solve for the thermodynamic potential in Eq.~(\ref{Omegaint}) together with the number equation $N=N_{\rm F}+2N_{\rm B}=-\partial \Omega/\partial\mu$. 

{\em Universal equation of state.} In experiment, the effective range $R_m$ is approximately a constant around resonance and furthermore, independent of magnetic quantum number $m$. As a result, we set $R_m=R$ in the following for simplicity. On the other hand, the resonances for $m=0$ ($z$-resonance) and $m=\pm 1$ ($xy$-resonance) are split due to magnetic dipole-dipole interactions~\cite{Ticknor2004}. In the vicinity of the $xy$-resonance, there are two possible molecular states in the closed channel, while for the $z$-resonance, there is only one. In $^{40}$K experiments, these two resonances are well separated and do not interfere with each other which allows us to investigate the properties of the gas for the  $xy$- and $z$-resonance separately. 
\begin{figure}[ht]
\includegraphics[width=0.4\textwidth]{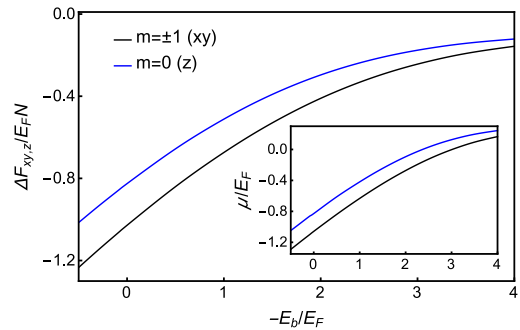}
\caption{Free energy $F_{xy,z}$ of the system as a function of $-E_b/E_F$ close to the $xy$- and $z$-resonances at $k_{\rm B}T=0.8 E_F$. In our calculation, we have set $k_FR=0.04$, appropriate to the experiment. $F_{xy}$ (black line) is always smaller than that of the $F_z$ (blue line) due to multiple molecular bound states. Inset shows the chemical potential $\mu$ as a function of $-E_b/E_F$ for the same set of parameters.}
\label{FigFreeE}
\end{figure}

The free energy of the system can be written in the following universal form around the $xy$- and $z$-resonances:  $F_{xy,z}=NE_Ff_{xy,z}(k_F R,-E_b/E_F,k_{\rm B}T/E_F)$. In this work, we focus on the dependences of $f_{xy,z}$ on $E_b/E_F$, assuming $k_FR=0.04$, a typical experimental value. Note that the scaling form for $F_{xy,z}$ works for both normal and superfluid phases. For normal state, we calculate the scaling functions $f_{xy,z}$ within NSR for $k_{\rm B}T=0.8E_F$ (see Figure \ref{FigFreeE}).  For both resonances, the free energy decreases monotonically from the BCS side to the BEC side. Throughout the crossover, $f_{xy}<f_z$ since there exist two molecular bound states for $xy$-resonance and this lowers its free energy. In fact, the difference $|f_{xy}-f_z|$ increases as one moves towards the BEC limit. Close to resonance, the reduction of free energy is of order of Fermi energy, indicating strong $p$-wave interactions.

\begin{figure}[ht]
\includegraphics[width=0.4\textwidth]{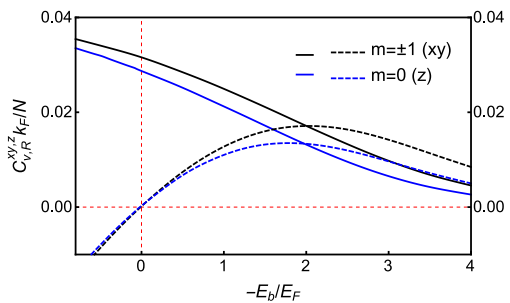}
\caption{Contacts $C_v^{xy,z}$ and $C_R^{xy,z}$ as a function of $-E_b/E_F$ for $k_{\rm B}T=0.8 E_F$ and $k_FR=0.04$. $C_v^{xy}$ (solid black line) and $C_v^{z}$ (solid blue line) decrease monotonically from the BEC to the BCS side with $C_v^{z}$ always smaller than $C_v^{xy}$. On the other hand, both $C_R^{xy}$ (dashed black line) and $C_R^{z}$ (dashed blue line) vanish at resonance $v=\pm\infty$ and depend on $-E_b/E_F$ non-monotonically. The magnitude of $C_R^{z}$ is always smaller than $C_R^{xy}$.} 
\label{FigContact}
\end{figure}

{\em P-wave contacts}. One of the most exciting aspects of the $p$-wave resonantly interacting Fermi gas is the existence of an extended set of universal relations involving the $p$-wave contacts. As in the $s$-wave case, the $p$-wave contacts determine the universal properties of the system including their response to external radio-frequency field, which has been utilized to measure the $p$-wave contacts in recent experiment~\cite{Luciuk2016}. Theoretically, one can calculate the values of the relevant contacts using the adiabatic theorems, which relate them to the variation of the free energy with respect to the scattering parameters:
\begin{equation}
\left.\frac{\partial F}{\partial {v_z^{-1}}}\right|_{R,T}=-\frac{\hbar^2C_{v}^z}{2M}; ~~\left.\frac{\partial F}{\partial {R^{-1}}}\right|_{v_z,T}=-\frac{\hbar^2C_{R}^z}{2M},
\end{equation}
and the similar equation for the $xy$-resonance. 

In Fig.~\ref{FigContact}, we show the calculated values of $C^{xy,z}_{v}$ and $C^{xy,z}_{R}$ within NSR for $k_FR=0.04$ and $k_{\rm B}T=0.8E_F$. $C^{xy,z}_{v}$ is monotonically decreasing as a function of $-E_b/E_F$ from the BCS to BEC side and is always positive, consistent with the general requirement~\cite{Yu2015}. $C_v^{xy}$ is always greater than $C_v^{z}$, indicating a  stronger dependences of $F_{xy}$ on $-E_b/E_F$ than $F_z$ due to multiple bound states in the closed channel. On the other hand, $C^{xy,z}_{R}$ shows non-monotonic behavior as a function of $-E_b/E_F$, with maximal value always achieved in the BCS side of the resonance. Note that even though $C^{xy,z}_{R}$ changes sign across the resonance and vanishes when $v=\pm\infty$, the magnitude of $C^{xy}_{R}$ is always larger than $C^{z}_{R}$, consistent with  stronger interactions around the $xy$-resonance.

{\em Superfluid transition temperatures}. Near a broad $s$-wave Feshbach resonance, superfluidity is the most robust at unitarity:  the coherence length is shortest and the critical current is largest~\cite{Zwerger2011}. Near a $p$-wave resonance, the pairing symmetry is richer~\cite{Ho2005,Cheng2005,Gurarie2005,Cheng2006,Gurarie2007}, and additional internal structure breaks scale invariance. In our formulation, apart from the splitting of resonance of $m=\pm 1$ ($xy$) and $m=0$ ($z$), the vertex function $\Gamma_m$ is diagonal in $m$. This means that close to $xy$-resonance around $T_c$, the superfluid order parameter is of the axial form, with gap function $\Delta_{\bf k}$ having the symmetry $Y_{1\pm 1}(\hat{k})$ or their superposition. The detailed form cannot be obtained from our calculation within NSR. It is known that, however, the ground state order parameter should be of the pure $Y_{1\pm 1}(\hat{k})$ form~\cite{Ho2005}. Close to the $z$-resonance, the order parameter is the standard polar form, for which $\Delta_{\bf k}\propto k_z$. Using the Thouless criterion, we can write the equation for $T_c$ close to the $z$-resonance as $\Gamma^{-1}_{z}({\bf 0},0)=0$, or explicitly
\begin{equation}
\frac{M^2\mu}{2\pi R}+\frac{M}{4\pi v}+\Pi^r_z({\bf 0},0)=0,
\label{Tc}
\end{equation}
and similar equation equation determining the $T_c$ for $xy$-resonance by replacing $\Pi^r_z$ with $\Pi^r_{xy}$. 
\begin{figure}[ht]
\includegraphics[width=0.4\textwidth]{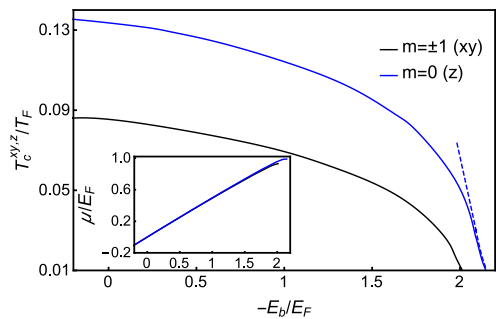}
\caption{Critical temperature $T_c$ as a function of $-E_b/E_F$ for the $xy$ (black line) and $z$-resonances (blue line). The dashed line is the asymptotic $T_c$ in the BCS limit for the $z$-resonance given by Eq.~\eqref{EqTcBCS}. Inset shows the respective chemical potentials for the $xy$- and $z$-resonances. } 
\label{FigTc}
\end{figure}

In Fig.~\ref{FigTc}, we show the calculated critical temperature for the $xy$- and $z$-resonance. For the $xy$-resonance, the critical temperature is always lower than that for the $z$-resonance. This is because for the $xy$-resonance, there are two molecular states in the closed channel which reduces the quantum degeneracy of the system. This is particularly evident in the BEC limit where all $N/2$ bosons are divided between two molecular states (with $m\pm 1$, each with number $N/4$ and density $n/4$), so the corresponding critical temperature for BEC is given by $T^{xy}_c/T_F=0.086$. For the $z$-resonance, there is only one molecular state with $m=0$ and the corresponding number density is $N/2V=n/2$. This gives a higher critical temperature $T^z_c/T_F=0.137$. These values are also consistent with the calculated $T_c$ via NSR in the BEC side. The general behavior of $T_c$ in our calculation is consistent with that of Ref.~\cite{Ohashi2005}, where, however, the dependence of $T_c$ on the low-energy scattering parameters $\{v_m,R_m\}$ is implicit since it uses the bare coupling constants with a cutoff. 

In the BCS limit, $v\to 0^-$, the attractive interaction is very weak and $T_c$ becomes very small. In this limit, Eq.~(\ref{Tc}) simplifies considerably and one can obtain the asymptotic value of $T_c$ as
\begin{equation}
T_c=\frac{8\gamma\tilde{\mu}}{\pi}T_F\exp\left[-\frac{8}{3}\tilde{\mu}^{\frac{3}{2}}\right]\exp\left[\frac{\pi\tilde{\mu}}{2k_FR}+\frac{\pi}{2k_F^3v}\right],
\label{EqTcBCS}
\end{equation}
where $\gamma=1.78107$ is the exponential of the Euler constant and $\tilde{\mu}=\mu/E_F$. Note the standard exponential dependences on the $p$-wave interaction parameter, the scattering volume $v$, similar to the $s$-wave case.  In addition, it should be noted that the dependence on the effective range $R$ is also non-analytic, indicating the importance of finite range corrections in a dilute $p$-wave superfluid. Setting $\tilde{\mu}=1$ in Eq.~(\ref{EqTcBCS}), one can equivalently write~\cite{Iskin2006}
\begin{equation}
T_c=\frac{8\gamma}{\pi}T_F\exp\left[-\frac{8}{3}\right]\exp\left[\frac{\pi}{2k_FR}\left(1-\frac{E_b}{2E_F}\right)\right],
\end{equation}
showing that $T_c$ changes rapidly around $E_b/E_F\approx 2$ when the quasi-bound $p$-wave state moves out of the scattering continuum.

At the critical temperature $T=T_c$, the corresponding chemical potential $\mu$ as a function of $E_b/E_F$ is shown in the inset of Fig.~\ref{FigTc}. The chemical potential $\mu$ is approximately half of the binding energy $E_b$ and is thus linear in $E_b$ over a wide range of interaction, but it quickly approaches $E_F$ around $E_b/E_F\approx 2$ in the BCS limit. When $\mu=0$, there is a very simple relation between the critical temperature and the scattering volume $v$ 
\begin{equation}
\left.\frac{1}{k_F^3v}\right|_{\mu=0}=\frac{2-\sqrt{2}}{2\sqrt{\pi}}\zeta({3}/{2})\left(\frac{T_c}{E_F}\right)^{3/2},
\label{Equ0}
\end{equation}
where $T_c$ is the critical temperature at $\mu=0$. We note that this relation is {\em independent} of the effective range $R$.

{\em Conclusion}. In this Letter, we studied the normal state properties of a resonantly interacting $p$-wave Fermi gas. The universal equation of state, the $p$-wave contacts and the superfluid transition temperatures $T_c$ are obtained using a two-channel formulation. We show that for the resonance with $m=\pm 1$, our estimation of the transition temperatures using actual experimental parameters is quite encouraging experimentally. There remain further important theoretical questions to be investigated such as the analogous Gor'kov-Melik-Barkhudarov correction to $T_c$ from medium polarizations~\cite{Gorkov1961}.

{\em Acknowledgement}. We thank Boyang Liu, Shanshan Ding and Joseph Thywissen for helpful discussions. This work is supported by Hong Kong Research Grants Council, General Research Fund, HKU 17306414, CRF, HKUST3/CRF/13G, and the Croucher Foundation under the Croucher Innovation Award.

\end{document}